\documentclass{article}
\usepackage{spconf,amsmath,graphicx}
\usepackage{multirow}
\usepackage{url}

\title{Self-supervised audio-visual speech representations learning by multimodal self-distillation}
\name{Jing-Xuan Zhang$^{1, 2}$, Genshun Wan$^{1, 2}$, Zhen-Hua Ling$^1$, Jia Pan$^2$, Jianqing Gao$^2$, Cong Liu$^2$}
\address{
$^1$University of Science and Technology of China, China \\
$^2$iFLYTEK Research, iFLYTEK Co. Ltd., China \\
\small{\texttt{\{nosisi, gswan\}@mail.ustc.edu.cn, zhling@ustc.edu.cn, \{jiapan, jqgao, congliu2\}@iflytek.com}}}

\begin{document}
\ninept
\maketitle
\begin{abstract}
In this work, we present a novel method, named AV2vec, for learning audio-visual speech representations by multimodal self-distillation. AV2vec has a student and a teacher module, in which the student performs a masked latent feature regression task using the multimodal target features generated online by the teacher. 
The parameters of the teacher model are a momentum update of the student. 
Since our target features are generated online, 
AV2vec needs no iteration step like AV-HuBERT and
the total training time cost is reduced to less than
one-fifth.
We further propose AV2vec-MLM in this study, which augments AV2vec with a masked language model (MLM)-style loss using multitask learning.
Our experimental results show that AV2vec achieved comparable performance to the AV-HuBERT baseline. When combined with an MLM-style loss, AV2vec-MLM outperformed baselines and achieved the best performance on the downstream tasks.
\end{abstract}
\begin{keywords}
self-supervised learning, audio-visual speech recognition, lipreading
\end{keywords}
\section{Introduction}
\label{sec:intro}
Humans are known to rely on both audio and visual information to perceive speech~\cite{mcgurk1976hearing}. 
Presented with corrupted and entangled sounds, the human perceptual system draws heavily on visual information to reduce ambiguities in audio~\cite{rahne2007visual}. In machine learning, a rich body of research has explored learning from the multi-modal audio-visual speech data. 
For instance, visual speech recognition (VSR), also known as lipreading, aims at predicting speech content with only visual input~
\cite{wand2016lipreading,martinez2020lipreading,Themos2017combine,Zhang_Richmond_Ling_Dai_2021}. Audio-visual speech recognition (AVSR) adopts both audio and visual modality input for speech recognition~\cite{mroueh2015deep,petridis2018audio,afouras2018deep,ma2021end,makino2019recurrent}. Because the visual signal is invariant to the acoustic interference, VSR and AVSR have wide applications such as communication in noisy environments or improved hearing aids.

Recent years have witnessed rapid development in self-supervised learning and it has achieved success in both computer vision~\cite{he2020momentum,chen2020simple,grill2020bootstrap,bao2022beit,he2022masked} 
and speech processing domain~\cite{schneider2019wav2vec,baevski2020wAV2vec,hsu2021hubert,DBLP:conf/icml/BaevskiHXBGA22}. 
Different from supervised learning, where labels provide an explicit discrimination task for learning, self-supervised learning methods employ pretext tasks for model pretraining.  
It enables to learn general representations from rich unlabeled data and to finetune the model on labeled data.
For audio and visual representations learning, previous studies leverage the natural synergy between the audio and visual channels of the video for cross-modal self-supervision. AVTS model proposes to learn general audio-visual representations via self-supervised temporal synchronization and a contrastive loss~\cite{korbar2018cooperative}. XDC leverages cross-modal audio-visual clustering for self-supervised learning~\cite{alwassel2020self}. A contrastive learning method for both global and local audio-visual representations has also been proposed~\cite{NEURIPS2021_38ef4b66}. Particularly, a newly proposed method by Shi et al. extends HuBERT~\cite{hsu2021hubert} from speech to audio-visual speech data~\cite{DBLP:conf/iclr/ShiHLM22}. AV-HuBERT 
 is trained iteratively by alternating between an offline feature clustering step and a masked prediction step until the model performance is no longer improved. It has achieved state-of-the-art performance on downstream tasks such as VSR~\cite{DBLP:conf/iclr/ShiHLM22} and AVSR~\cite{DBLP:journals/corr/abs-2201-01763}.

Despite its impressive performance, pretraining an AV-HuBERT model is time-consuming because of multiple iterations. 
In this work, we propose AV2vec, a self-supervised model for learning audio-visual representations without iterative training.
AV2vec employs a student and a teacher module. The student performs a masked latent feature regression task, in which the targets are generated online by the teacher module. And the parameters of the teacher are a momentum update of the student. For the student, we apply three types of corruption for multimodal input. First, the audio signal input is corrupted by adding noise randomly. Second, we apply modality-independent span masks for intermediate hidden features. Third, a modality dropout is employed, namely that either audio or video features are dropped randomly. For the teacher, it always accepts full clean audio and visual input. 
Since our method generates targets online without iterations, AV2vec can be trained much faster than AV-HuBERT. Moreover, motivated by the success of AV-HuBERT~\cite{DBLP:conf/iclr/ShiHLM22} and HuBERT~\cite{hsu2021hubert}, this work further presents AV2vec-MLM that augments AV2vec with a masked language model (MLM)-style loss using multitask learning.

In our experiments, we compared the performance of our proposed AV2vec method with the AV-HuBERT baseline comprehensively on the ASR, VSR, and AVSR downstream tasks. For the visual stream, two types of input were investigated including the entire face and the traditional lip region-of-interest (ROI). Our experimental results suggest that AV2vec achieved comparable performance to the AV-HuBERT baseline. When AV2vec is augmented with an MLM-style loss, AV2vec-MLM achieved the best results on downstream tasks.


\begin{figure*}[t]
    \centering
    \includegraphics[width=0.95\textwidth]{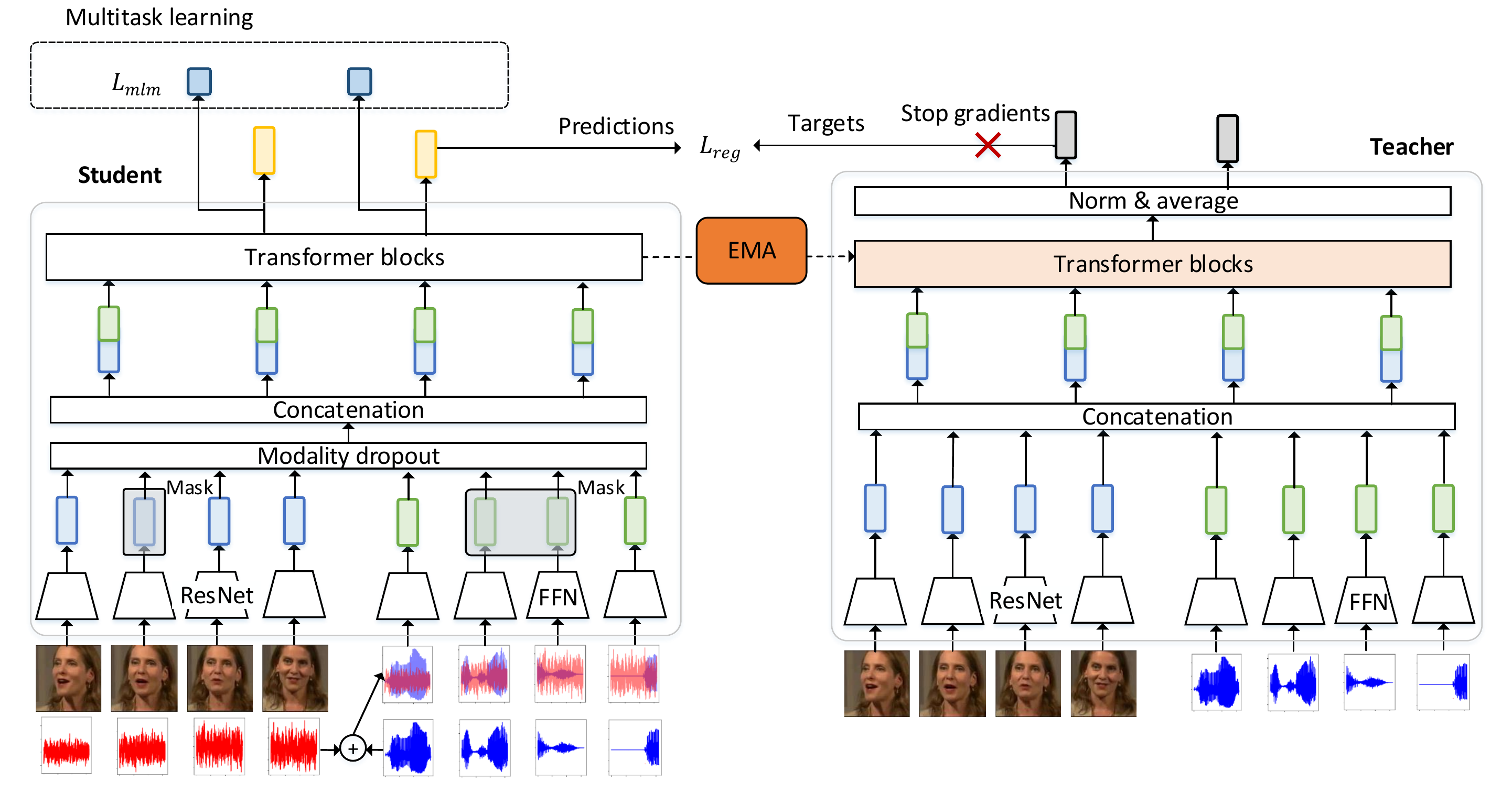}
    \caption{Illustration of our proposed AV2vec model. AV2vec contains a student and a teacher module. The student performs a masked latent feature regression task and the targets are generated online by the teacher. The teacher weights are an exponentially moving average (EMA) of the student parameters.}
    \label{fig:figure1}
\end{figure*}

\section{Related Works}
\label{sec:related}

Many semi-supervised and self-supervised learning methods employ a momentum teacher for generating continuously refined targets online~\cite{DBLP:conf/nips/TarvainenV17,DBLP:conf/interspeech/HiguchiMRH21,grill2020bootstrap,caron2021emerging,zhou2022image,DBLP:conf/icml/BaevskiHXBGA22}.
Mean teacher~\cite{DBLP:conf/nips/TarvainenV17} penalizes a consistency cost between the student and the exponentially moving average (EMA) based teacher for semi-supervised learning.
MPL~\cite{DBLP:conf/interspeech/HiguchiMRH21} adopts a momentum teacher to generate pseudo-labels for semi-supervised ASR. 
For self-supervised learning, BYOL~\cite{grill2020bootstrap} relies on an online and target network, where the online network is trained to predict the target network representations of the same image under a different augmented view. 
DINO~\cite{caron2021emerging} adopts a similar method to BYOL and interprets it as a form of self-knowledge distillation with no labels.
Our work is inspired by data2vec~\cite{DBLP:conf/icml/BaevskiHXBGA22}, which predicts latent representations of the full input based on a masked view of the input. Despite that data2vec is a general framework for different modality inputs, it's not designed for learning representations from multimodal data within a signal model. 
Different from data2vec, our AV2vec model learns from audio-visual multimodal data and enables cross-modal knowledge distillation.
Also,
a novel multitask learning
method that combines AV2vec with an MLM-style loss
is presented in this work.


\section{Proposed Method}
\label{sec:proposed}

Our method consists of a student and a teacher module, where the student is trained with a masked latent feature regression task and the
targets are generated online by the teacher. The model is illustrated in \figurename~\ref{fig:figure1}.

\subsection{Input of the model}
For the audio stream, Mel filter banks feature is used as input. For the visual stream, two types of input are compared including the conventional lip region-of-interest (ROI) and the entire face images.
Our previous study~\cite{zhang2022lip} suggests the entire faces contain more detailed information than lips which may benefit visual representation learning. And the experimental results have shown that adopting faces as visual input led to an improved lipreading accuracy.

\subsection{Student module}
\label{subsec:student}
Let the clean audio be $\boldsymbol{A}^c =[\boldsymbol{a}^c_1, \boldsymbol{a}^c_2,\dots, \boldsymbol{a}^c_T]$, the video input be $\boldsymbol{V} =
[\boldsymbol{v}_1,\boldsymbol{v}_2,\dots,\boldsymbol{v}_T]$. 
In order to increase our model's robustness to acoustic noise, the audio input
of the student $\boldsymbol{A} =[\boldsymbol{a}_1, \boldsymbol{a}_2,\dots, \boldsymbol{a}_T]$  is obtained by adding noise to clean audio randomly  with a probability $p_{noise}$. The audio and visual input is processed by modality-specific feature extractors for producing
audio intermediate feature $\boldsymbol{F}^a = [\boldsymbol{f}^a_1, \boldsymbol{f}^a_2,\dots,\boldsymbol{f}^a_T]$ and visual intermediate feature
 $\boldsymbol{F}^v = [\boldsymbol{f}^v_1, \boldsymbol{f}^v_2,\dots,\boldsymbol{f}^v_T]$ respectively. A feed forward network and a 3DCNN-ResNet~\cite{afouras2018deep} network are adopted for audio and visual feature extractor respectively. Then features of each modality are corrupted independently by span-based masking. Specifically, 
 let $M_a$ be a set of masked indices for audio, $\tilde{\boldsymbol{F}}^a
 = r(\boldsymbol{F}^a, M_a)$ denotes audio feature $\boldsymbol{F}^a$
 is corrupted by replacing $\boldsymbol{f}^a_t$ with an embedding $\boldsymbol{e}^a$ if $t \in M_a$. For corrupting visual features, the same strategy as the audio is used but with different mask indices
 $M_v$ and a different embedding $\boldsymbol{e}^v$. We also tried to corrupt audio and visual features with 
 the same mask indices in our preliminary experiments, while the results 
 were worse than the modality-independent masking. Then a modality dropout similar to that in AV-HuBERT~\cite{DBLP:conf/iclr/ShiHLM22} is employed to train the student 
 in absence of audio or video stream randomly. Specifically, with a probability
 $p_m$, both modalities are used. The probability of selecting audio is
 $p_a$ when only one modality is used, and features of the other 
 modality are set to all zeros.  Audio-visual representations
 are then obtained as $\boldsymbol{F}^{av} = Concat(\tilde{\boldsymbol{F}}^a, \tilde{\boldsymbol{F}}^v)$,
 where $Concat( \cdot )$ denotes channel-wise concatenation.
 $\boldsymbol{F}^{av}$ is fed into a stack of Transformer blocks for generating
multimodal contextualized features, which are passed through
a linear projection layer for regressing to latent representations generated by the teacher. A mean square error loss is used as 
\begin{equation}
L_{reg} = \sum_{t \in M_a \cup M_v} || \boldsymbol{x}_t - \boldsymbol{y}_t||^2,
\end{equation}
where $\boldsymbol{x}_t$ is the prediction and $\boldsymbol{y}_t$ is the target produced by the teacher. Notes that we only penalized the predictions in masked regions.

\subsection{Teacher module}
\label{subsec:teacher}
The teacher module consumes both clean audio $\textbf{A}^c$ and
video $\textbf{V}$ as input. Intermediate feature masking or modality
dropout is not applied for the teacher. Therefore, it generates
hidden representations with full context and multimodal input.
The teacher weights are updated by an exponentially moving 
average (EMA) of the student as
\begin{equation}
\boldsymbol{\theta}_i = \lambda * \boldsymbol{\theta}_{i-1} + (1-\lambda) * \boldsymbol{\phi}_{i},
\end{equation}
where $\boldsymbol{\theta}$ and $\boldsymbol{\phi}$ denote the teacher and the student parameters, $i$ denotes the update step and $\lambda$ is a hyper-parameter controls the update step.
$\lambda$ is linearly increased from $\lambda_b$ to $\lambda_e$
at the initial $n$ updates and is kept constant in the rest
training steps.
This momentum update is 
applied only for the Transformer blocks and the teacher's feature extractors share
the parameters with the student. No gradients flow to the teacher 
and only the student is trained with the back-propagation algorithm.
For obtaining targets of the student, instance normalization~\cite{DBLP:journals/corr/UlyanovVL16} is applied for 
the last $k$ Transformer blocks' hidden outputs and then they are
averaged following data2vec~\cite{DBLP:conf/icml/BaevskiHXBGA22}.

\subsection{Multitask learning}
\label{subsec:multi}
In this section, we further propose to augment AV2vec with a  
masked language model (MLM)-style loss like that of HuBERT, which is denoted as
AV2vec-MLM. It employs a separate linear projection layer
 on the top of the student's Transformer blocks
for discrete targets classification, which is penalized with a loss as 
\begin{equation}
L_{mlm} = \sum_{t \in M_a \cup M_v} \text{CE}(l_t, \boldsymbol{p}_t),
\end{equation}
where $l$ and $\boldsymbol{p}_t$ denote the category of the discrete target and the predicted probability distribution respectively, and $\text{CE}(\cdot)$ denotes
the cross entropy loss function.
The targets can be derived from hidden features of a pretrained AV2vec
or AV-HuBERT model by k-means clustering.
AV2vec-MLM is optimized with multitask learning and the total training loss is a summation of the latent
feature regression loss and the MLM-style loss as
\begin{equation}
L_{MT} = L_{reg} + L_{mlm}.
\end{equation}
MLM-style loss serves as an additional supervision signal  and is expected to train the model more efficiently. However, AV2vec-MLM requires a pretrained model for generating targets of MLM-style loss therefore it costs more training time than AV2vec. Hence, it provides a trade-off between the training cost and the model performance.

\section{Experiments}
\subsection{Implementation details}
Experiments were conducted based on the LRS3 dataset~\cite{Afouras18d}, which contains
over 400 hours of English videos. 
All training data was used for pretraining and
a subset containing 30 hours of training data
or the full training data was used for finetuning.
MUSAN dataset~\cite{Snyder2015MUSANAM} was used
and we followed the protocol of \cite{DBLP:journals/corr/abs-2201-01763}
for sampling and adding noise to the audio input
of the student and the test set.
We constructed the test sets at
five SNR levels from -10 dB to 10 dB with a 5 dB interval.
For extracting lip ROI and the entire face
 images as visual input, we followed our previous study~\cite{zhang2022lip}.

For the network structure, AV2vec adopted the same configuration
as the ``BASE" version of AV-HuBERT~\cite{DBLP:conf/iclr/ShiHLM22} to achieve the comparable
computation complexity and number of parameters to it.
A single feed forward layer and 3DCNN-ResNet are used for audio and
visual feature extractor.
It has 12 Transformer blocks and the embedding dimension/feed forward dimension /attention heads were 768/3072/12 respectively. 
$p_{noise}$ was set to 0.25. 
80\% of audio features
and 30\% of video features were masked for the student during pretraining. $p_m$ and
$p_a$ were set to 0.5 and 0.5 respectively.
$\lambda_b$, $\lambda_e$, and $n$ were set to 0.999, 0.9999, and 30k respectively.
By default, we averaged representations of the last 8 layers for the teacher following~\cite{DBLP:conf/icml/BaevskiHXBGA22}.
Adam optimizer was used with a peak learning rate of $5 \times 10^{-4}$. The learning rate was first warmup linearly for the first 3\% of updates and then kept constant 
for 90\% of updates followed by exponentially 
decay over the remaining 7\%.

After pretraining, a 6-layer Transformer
decoder was used and the pretrained model was adopted as the encoder. 
During finetuning, the encoder was frozen at the initial $i$ steps and then jointly finetuned with the decoder, where $i$ was determined using the validation set. Our methods
were evaluated on three types of audio-visual downstream tasks, including 
ASR, VSR, and AVSR. 
The noise was added during both pretraining and
finetuning stages except finetuning for VSR task.

Three methods were compared in our experiments and they are described
as follows:

\textbf{AV-HuBERT~\cite{DBLP:conf/iclr/ShiHLM22,DBLP:journals/corr/abs-2201-01763}:} The baseline method. We added noise to the student audio input
during both pretraining and finetuning stage following~\cite{DBLP:journals/corr/abs-2201-01763}. The model was
constructed based on the 4th iteration checkpoint\footnote{\url{https://facebookresearch.github.io/av_hubert/}} publicly released by the author to save the training time. 
 
\textbf{AV2vec:} Our proposed method which is described in Section~\ref{sec:proposed}.

\textbf{AV2vec-MLM:} Our proposed method which is augmented with an MLM-style loss as described in Section~\ref{subsec:multi}. The discrete labels were derived from 
the pretrained AV-HuBERT of the 4th iteration to be directly comparable with the baseline.

\subsection{Training costs}
AV2vec and one iteration of AV-HuBERT both cost around
8 days of training time for 400k updates on 8 Tesla A100 GPU cards. Therefore, training AV-HuBERT
 for 5 iterations from MFCC features following ~\cite{DBLP:conf/iclr/ShiHLM22} will spend around 5-6 times as long as training AV2vec if taking also into account the time costs of the clustering step. Training AV2vec-MLM from scratch requires
 more training time than AV2vec, since it relies on a pretrained model to generate
 discrete targets.
 Despite that, AV2vec-MLM still costs less
 than a half of training time than AV-HuBERT
 if it employs AV2vec for producing targets of MLM loss.
Overall, our proposed method yields a significantly better training efficiency
 than AV-HuBERT does.
 
\subsection{Main results}

\begin{table*}[]
     \caption{Word error rate (WER \%) results on the LRS3 test set.``Hrs'' denotes
     the amount of labeled training data. ``Vt'' denotes the visual type for visual stream input, i.e. lips or faces. `` $\infty$'' denotes clean audio. 
     }
     \label{tab:tab1}
    \centering
    
    \begin{tabular}{c  c  c  |  c  | c   c   c  c  c   c |   c  c  c  c  c  c }
    \hline
    \hline
    Hrs & Method & Vt &  VSR & \multicolumn{6}{c | }{ASR, SNR (dB)=} & \multicolumn{6}{c}{AVSR, SNR (dB)=} \\
    & & & & -10 & -5 & 0 & 5 & 10 & $\infty$ & -10 & -5 & 0 & 5 & 10 & $\infty$ \\
    \hline
    1.4k & Afouras et al.~\cite{Afouras18d} & lip & 58.9 & -- & -- & -- & -- & -- & 8.3 & -- & -- & -- & -- & -- & 7.2 \\ 
    595  & Xu et al.~\cite{xu2020discriminative} & lip      & 57.8 & -- & -- & -- & -- & -- & 7.2 & -- & -- & -- & -- & -- & 6.8 \\
    31k  & Makino et al.~\cite{makino2019recurrent} & lip  & 33.6 & -- & -- & -- & -- & -- & 4.8 & -- & -- & -- & -- & -- & 4.5 \\
    \hline
    \multirow{5}{*}{30}   & AV-HuBERT & lip & 47.1 & -- & -- & -- & -- & -- &  -- & 30.2 & 18.3 & 10.7 & 7.4 & \textbf{6.3} & \textbf{5.4} \\
    & AV-HuBERT & face & 42.5 & 81.5 &  52.8 & 24.7 &  13.2 & \textbf{8.5} & 5.7 & 28.0 & 17.4 & 10.9 & 7.5 & 6.4 & \textbf{5.4} \\
    & AV2vec    & lip  & 45.1 & -- & -- & -- & -- & -- & -- & 30.2 & 18.6 & 10.9 &  7.9 & 6.8 & 5.8 \\
    & AV2vec    & face & \textbf{39.4} & 81.6 & 53.1 & 25.1 & 12.6 & 8.8 & 6.1 & 26.9 & 17.2 & 10.5 & 
         7.8 & 6.8 & 5.6 \\
    & AV2vec-MLM  & face & \textbf{39.4} & \textbf{81.3} & \textbf{51.7} & \textbf{23.9} & 
         \textbf{12.4} & \textbf{8.5} & \textbf{5.6} & \textbf{26.8} & \textbf{16.3} & \textbf{10.0} & \textbf{7.3} & \textbf{6.3} & \textbf{5.4} \\
    \hline
      \multirow{5}{*}{433}   & AV-HuBERT & lip & 40.3 &  -- & -- & -- & -- & -- &  -- & 27.5 & 15.1 & 7.1 & 4.5 & 3.4 & 2.6 \\
    & AV-HuBERT & face & 34.8 & 71.6 & 46.7 & 19.8 & \textbf{8.9} & 5.5 & 2.8 & 25.0 & 14.2 & 7.0 & 4.3 & 3.5 & 2.7 \\
    & AV2vec    & lip  & 39.9 & -- & -- & -- & -- & -- & -- & 26.8 & 14.7 & 7.5 &  4.4 & 3.7 & 2.6 \\
    & AV2vec    & face & \textbf{34.3} & 72.6 & 48.1 & 20.7 & 9.3 & 6.1 &  2.9 & 23.9 & 13.5 & \textbf{6.7} & 
         4.3 & 3.5 & 2.8 \\
    & AV2vec-MLM  & face & 34.4 & \textbf{71.3} & \textbf{46.3} & \textbf{19.5} & 
         \textbf{8.9} & \textbf{5.2} & \textbf{2.7} & \textbf{23.4} & \textbf{13.3} & \textbf{6.7} & \textbf{4.2} & \textbf{3.2} & \textbf{2.5} \\
    \hline
    \hline
    \end{tabular}
    \label{tab:my_label}
\end{table*}

Our experimental results are summarized 
in \tablename~\ref{tab:tab1}. 
The pretrained models were finetuned and evaluated in VSR, ASR, and AVSR 
tasks. 
In VSR task, it's observed that the face-based model outperformed the
lip-based counterpart consistently for both AV-HuBERT and our proposed method. Compared with the AV-HuBERT baseline, our method
obtained better lipreading accuracy. AV2vec and AV2vec-MLM achieved
close performance in this task.
In ASR task, the face-based pretrained models were adopted for finetuning. We observed that AV2vec achieved 
close but slightly worse results than the AV-HuBERT baseline.
When AV2vec was augmented with an MLM-style loss, ASR performance
can be further improved and AV2vec-MLM achieved the lowest WER. 
For AVSR task, we observed that the face based model achieved 
better performance than the lip based counterpart under low SNR ($<$0dB)
conditions while they achieved close performance under high SNR ($>$0dB) conditions.
These results also suggest that faces provide better visual information than lips for speech recognition because the AVSR model relies more heavily on visual information
under a higher noise condition.
In general, AV2vec achieved
better performance under low SNR conditions and
worse performance under high SNR conditions compared with the baseline.
AV2vec-MLM model achieved the best performance in AVSR task at most SNR levels and this proves the effectiveness of multitask learning for AV2vec for generating better audio-visual representations. We also compared our method with previous supervised-based models (\tablename~\ref{tab:tab1}, row 2-4) and the results
suggest our method achieved better or competitive performance in three tasks with less training data. 

\begin{table}[t]
  \caption{WER (\%) results on the LRS3 test set in ablation study of modality dropout. ``-m" and ``-x'' denote
  dropout the same and the opposite modality for the teacher respectively if
  one modality is dropped for the student.}
    \label{tab:tab2}
    \centering
    \begin{tabular}{c | c  c c  c  c  c}
    \hline
    \hline
    \multirow{2}{*}{Method} & \multicolumn{6}{c}{SNR (dB)=} \\
    
    & -10 & -5 & 0  & 5 & 10  & $\infty$ \\
    \hline
    AV2vec & \textbf{26.9} & \textbf{17.2} & \textbf{10.5} & \textbf{7.8} & \textbf{6.8} & \textbf{5.6}  \\
    AV2vec-m & 37.3 & 26.0 & 17.1 & 13.4 & 11.7 & 9.9  \\
    AV2vec-x & 37.8 & 23.6 & 15.1 & 11.6 & 10.1  & 8.9 \\ 
     \hline
    \hline
    \end{tabular}
  
\end{table}

\subsection{Ablation studies}
In ablation studies, we investigated the modality dropout used by the teacher. Three configurations were compared including our proposed method \textbf{AV2vec} which adopts no teacher modality dropout,
\textbf{AV2vec-m} in which the teacher dropouts the same modality as the student and \textbf{AV2vec-x} 
dropouts the opposite modality if one modality is dropped in the student.
 Ablation studies were conducted based
on the 30-hour labeled data for the AVSR task and results are presented in \tablename~\ref{tab:tab2}. From the table, our proposed method achieved the
best results, which confirmed the effectiveness of multimodal distillation
used in our proposed method.

Ablation studies that averaged different numbers of teacher layer
representations were also conducted and the results are shown 
in \tablename~\ref{tab:tab3}. From the table, it's observed that multiple
layers improved AVSR accuracy, which is consistent with the previous study~\cite{DBLP:conf/icml/BaevskiHXBGA22}. Our method averaging the last 8 layers following~\cite{DBLP:conf/icml/BaevskiHXBGA22} achieved
the best performance.

\begin{table}[t]
  \caption{WER (\%) results on the LRS3 test set when averaging the last $k$ layer teacher
  hidden representations for generating the student's target. $^\dagger$ denotes the proposed method.}
    \label{tab:tab3}
    \centering
    \begin{tabular}{c | c  c  c  c  c  c}
    \hline
    \hline
    \multirow{2}{*}{ $k$ } & \multicolumn{6}{c}{SNR (dB)=} \\
    
    & -10 & -5 & 0  & 5  & 10 &  $\infty$ \\
    \hline
    12 & 30.6 & 20.1 & 12.6 & 9.9 & 8.9 & 7.7 \\
    8$^\dagger$ & \textbf{26.9} & \textbf{17.2} & \textbf{10.5} & \textbf{7.8} & \textbf{6.8} & \textbf{5.6}  \\
    4 & 32.7 & 20.7 & 13.0 & 9.5 & 7.9 & 6.7  \\
    1 & 62.7 & 55.9 & 50.0 & 47.9 & 46.7  & 46.1 \\ 
    \hline
    \hline
    \end{tabular}
  
\end{table}

\section{Conclusion}
In this work, we propose AV2vec, a self-supervised training method
for audio-visual speech representation learning. It has a student 
and a teacher module, where the student performs a masked latent feature
regression task and the multimodal targets are generated online by the momentum-based teacher.
Since our method is trained without iteration, it reduces training time significantly
compared to AV-HuBERT.
We also present AV2vec-MLM, in which AV2vec is augmented with an MLM-style loss
 by multitask learning. 
 Our experimental results indicate that AV2vec learned better visual representations but slightly worse audio representations compared to the Av-HuBERT baseline.
 Combining AV2vec with an MLM-style loss further improved the performance and our proposed AV2vec-MLM method achieved the best results on downstream tasks including ASR, VSR, and AVSR.



\bibliographystyle{IEEEbib}
\bibliography{strings,refs}

\end{document}